\documentclass[draftclsnofoot, onecolumn, 12pt]{IEEEtran}
\IEEEoverridecommandlockouts

\usepackage{fancyhdr}
\usepackage{setspace}
\usepackage{ifpdf}

\usepackage{cite}

\ifCLASSINFOpdf
  \usepackage[pdftex]{graphicx}
  \graphicspath{{./figs/}{./figs/}{./figs/}}
  \DeclareGraphicsExtensions{.pdf,.jpeg,.png}
\else
  \usepackage[dvips]{graphicx}
  \graphicspath{{./figs/}}
  \DeclareGraphicsExtensions{.eps}
\fi

\usepackage[cmex10]{amsmath}
\usepackage{amsfonts}
\usepackage{dsfont}

\usepackage{algorithmic}
\usepackage[ruled]{algorithm2e}

\usepackage{array}

\usepackage[tight,footnotesize]{subfigure}

\usepackage{enumerate}

\usepackage{url}

\usepackage{pict2e}
\usepackage{color}
\usepackage{transparent}
\usepackage{booktabs}
\usepackage{multirow}

\usepackage{acronym}
\acrodef{IA}{interference alignment}
\acrodef{PRPC}{per-\acp{RRU} power constraint}
\acrodef{PAPC-Inequality}{\acp{PAPC} where the constraints hold with inequality}
\acrodef{PAPC-Equality}{\acp{PAPC} where the constraints hold with equality}
\acrodef{PRSPC}{per-\ac{RRU} sum power constraint}
\acrodef{PRPEC}{per-\ac{RRU} power equality constraint}
\acrodef{SNR}{signal to noise ratio}
\acrodef{SINR}{signal to noise plus interference ratio}
\acrodef{BF}{beamforming}
\acrodef{AMIL}{alternating minimization of interference leakage}
\acrodef{IL}{interference leakage}
\acrodef{CDF}{cumulative distribution function}
\acrodef{PDF}{probability density function}
\acrodef{DAS}{distributed antenna systems}
\acrodef{CAS}{centralized antenna systems}
\acrodef{EGT}{equal gain transmission}
\acrodef{PAPR}{peak-to-average power ratio}
\acrodef{DoF}{degrees of freedom}
\acrodef{MIMO}{multiple-input, multiple-output}
\acrodef{IC}{interference channel}
\acrodef{ASP}{antenna-selective pathloss}
\acrodef{RRU}{remote radio unit}
\acrodef{SPC}{sum-power constraint}
\acrodef{PAPC}{per-antenna power constraint}
\acrodef{TDMA}{time-division multiple access}
\acrodef{ZF}{zero-forcing}
\acrodef{CSI}{channel state information}
\acrodef{SCM}{spatial channel model}
\acrodef{3GPP}{3rd Generation Partnership Project}

\newcommand{\norm}[1]{ \left\| {#1} \right\| }
\newcommand{\abs}[1]{ \left| {#1} \right| }

\newcommand{\chisqrcdf}[2]{ \mathcal{Q}\left( {#1}; {#2} \right) }

\newcommand{\rruset}{\mathcal{R}}

\newtheorem{theorem}{Theorem}
\newtheorem{proposition}{Proposition}

\def\bb0{{\mathbb{0}}}

\def\ba{{\mathbf{a}}}
\def\bb{{\mathbf{b}}}

\def\bd{{\mathbf{d}}}

\def\bff{{\mathbf{f}}}

\def\bm{{\mathbf{m}}}

\def\bx{{\mathbf{x}}}
\def\by{{\mathbf{y}}}
\def\bz{{\mathbf{z}}}
\def\b0{{\mathbf{0}}}

\def\bA{{\mathbf{A}}}

\def\bD{{\mathbf{D}}}

\def\bF{{\mathbf{F}}}

\def\bH{{\mathbf{H}}}
\def\bI{{\mathbf{I}}}
\def\bJ{{\mathbf{J}}}

\def\bM{{\mathbf{M}}}
\def\bN{{\mathbf{N}}}

\def\bR{{\mathbf{R}}}
\def\bS{{\mathbf{S}}}

\def\bW{{\mathbf{W}}}
\def\bX{{\mathbf{X}}}

\def\sf0{{\mathsf{0}}}

\def\Nt{{N_\mathrm{t}}}         %
\def\Nr{{N_\mathrm{r}}}         %
\def\Ns{{N_\mathrm{s}}}         %
\def\Nrru{{N_\mathrm{RRU}}}         %
\def\Rsum{\bar{R}_\mathrm{sum}}

\newtheorem{assumption}{Assumption}

\begin{document}

\title{Interference Alignment in Distributed Antenna Systems\thanks{This work was supported in part by the Army Research Laboratory under contract W911NF- 10-1-0420 and a gift from Huawei Technologies, Inc. This work has appeared in part at the 2011 IEEE International Symposium on Information Theory~\cite{starr2011interference}.}
}

\author{\IEEEauthorblockN{Jonathan~Starr, Omar~El~Ayach, and Robert~W.~Heath,~Jr.\footnote{The authors are with The University of Texas at Austin, Austin, TX 78712 USA (e-mail: {jstarr, oelayach, rheath}@utexas.edu).}}}

\maketitle

\begin{abstract}
Interference alignment (IA) is a cooperative transmission strategy that improves spectral efficiency in high signal-to-noise ratio (SNR) environments, yet performs poorly in low-SNR scenarios. This limits IA's utility in cellular systems as it is ineffective in improving cell-edge data rates. Modern cellular architectures such as distributed antenna systems (DAS), however, promise to boost cell-edge SNR, creating the environment needed to realize practical IA gains. Existing IA solutions cannot be applied to DAS as they neglect the per-remote-radio power constraints imposed on distributed precoders. This paper considers two types of distributed antenna IA systems: ones with a limit on maximum per-radio power, and ones with a strict equality constraint on per-radio power. The rate-loss incurred by a simple power back-off strategy, used in systems with maximum power constraints, is characterized analytically. It is also shown that enforcing strict power constraints avoids such a rate-loss but negatively affects IA feasibility. For such systems, an IA algorithm is proposed and feasibility conditions are derived based on the concept of system properness. Finally, numerical results validate the analysis and demonstrate that IA and DAS can be successfully combined to mitigate inter-cell interference and improve performance for most mobile users, especially those at the cell-edge.

\end{abstract}

\section{Introduction} \label{sec:introduction}

Interference alignment (IA) is a transmission strategy that is designed to maximize the number of non-interfering symbols that can be simultaneously communicated over an interference network \cite{cadambe2008interference}. By doing so, IA achieves the maximum degrees of freedom in a variety of single or multi-antenna interference channels and consequently allows systems to approach capacity in the high signal-to-noise ratio (SNR) regime. In the more practically relevant regime of low-to-medium SNR, however, IA's sum-rate performance has been shown to be suboptimal \cite{el2013practical, tresch2009cellular, tresch2009clustered, razaviyayn2010linear, shi2011iteratively}. This limits the applicability of IA in cellular systems since it does little to improve the rates of the disadvantaged cell-edge users that experience low SNRs~\cite{spencer2004introduction}. To overcome this practical shortcoming, prior work has focused on the development of enhanced precoding strategies that improve on IA's low-SNR performance. The unifying concept behind the algorithms in \cite{razaviyayn2010linear, shi2011iteratively, santamaria2010maximum, peters2011cooperative} is to potentially forgo perfect alignment and optimize objective functions that are more tightly related to system sum-rate. In addition to better algorithms, however, IA's utility can be improved by considering network architectures that may be more suitable for such a high-SNR transmission strategy. One such network is cellular systems with distributed antennas.

Distributed antenna systems (DAS) enhance the performance of cellular networks by supplementing or completely replacing traditional base stations, located at the cell-center, with a number of remote radio units (RRUs) positioned at distinct locations throughout the cell \cite{heath2013current, dai2005capacity}. The RRUs are then connected to each other via wired high-bandwidth low-latency links. Thus, the multi-RRU system acts as a single multi-antenna transmitter. By pushing antennas closer to the edge of the cell, DAS provide macrodiversity \cite{dai2005capacity, tse2005fundamentals} and improve the SNR experienced by cell-edge users or users in badly shadowed areas. Prior work on DAS has established the value of remote radios in improving coverage indoors \cite{chow1994performance, yanikomeroglu1993cdma}, reducing outages throughout the cell \cite{chen1996simulation, obaid2000reverse, kerpez1996radio}, increasing both uplink \cite{clark2001distributed} and downlink capacity \cite{choi2007downlink, zhuang2003spectral}, and improving fairness among users \cite{toufik2006wideband, zhu2010low}. The work in \cite{wang2009antenna, park2012antenna, park2011antenna} shows that even larger gains in outage or average capacity can be achieved by careful RRU placement. The work in \cite{dai2005capacity, chow1994performance, yanikomeroglu1993cdma, chen1996simulation, obaid2000reverse, kerpez1996radio, clark2001distributed, choi2007downlink, zhuang2003spectral, toufik2006wideband, zhu2010low, wang2009antenna, park2012antenna, park2011antenna}, however, was predominantly focused on single-antenna RRUs with simple point-to-point transmission strategies such as blanket transmission (all RRUs transmit at full power) or RRU selection (only one RRU is chosen for transmission/reception). Thus, the work in \cite{dai2005capacity, chow1994performance, yanikomeroglu1993cdma, chen1996simulation, obaid2000reverse, kerpez1996radio, clark2001distributed, choi2007downlink, zhuang2003spectral, toufik2006wideband, zhu2010low, wang2009antenna, park2012antenna, park2011antenna} does not capture the full potential of multi-antenna RRUs which
could enable enhanced transmission and interference management strategies.

To investigate the potential of multi-antenna RRUs, the work in \cite{xiao2003information, dai2006distributed, dai2007some, roh2003performance, feng2008size} considered the problem of trading off the \emph{number of RRUs} with the \emph{number of antennas per RRU} in cellular networks with constant antenna density. When single cells were considered \cite{xiao2003information, dai2006distributed, dai2007some}, maximizing macrodiversity via a denser deployment of single-antenna RRUs was shown to be optimal. For the more relevant multi-cell case, however, \cite{roh2003performance, feng2008size} showed that provisioning each RRU with multiple antennas yields large gains over both completely distributed and completely co-located configurations. The work in \cite{xiao2003information, dai2006distributed, dai2007some, roh2003performance, feng2008size}, however, is limited to non-cooperative single-user transmission strategies. DAS with multi-user transmission strategies were considered in~\cite{gan2007sum, dai2008distributed, feng2011deployment, heath2011multiuser, song2004downlink, wu2010radio} and their performance was shown to be promising on both the uplink \cite{gan2007sum, dai2008distributed, feng2011deployment} and downlink \cite{heath2011multiuser, song2004downlink, wu2010radio}. The work on multi-user support in DAS, however, often adopts a subset of the following non-trivial simplifications: (i) considering simplified multi-user access strategies\cite{wu2010radio}, (ii) limiting analysis to the more tractable uplink system~\cite{gan2007sum, dai2008distributed, feng2011deployment}, (ii) neglecting the per-RRU power constraints imposed on DAS~\cite{song2004downlink, heath2011multiuser}, (iii) treating intra-cell interference as noise~\cite{wu2010radio, song2004downlink}, and (iv) neglecting out-of-cell interference~\cite{gan2007sum, dai2008distributed, feng2011deployment, song2004downlink}.

In this paper we consider the application of interference alignment to the downlink of multi-cell systems with distributed antennas. The motivation for combining IA with DAS is two-fold. First, DAS may help overcome or avoid IA's low-SNR weakness by boosting signal power at the cell-edge. Second, IA may constitute an effective inter-cell interference management strategy for multi-cell DAS. Since existing IA solutions neglect per-RRU (or per-antenna) power constraints, and are thus not applicable to DAS, we focus on reevaluating the possibility and performance of IA in these more tightly constrained systems. We note that while the antenna's geographic separation in DAS necessitates the consideration of such power constraints, per-antenna constraints are in fact of practical relevance in all wireless transceivers that must operate with power-efficiency in mind, i.e., even those with co-located antennas \cite{love2003equal, sesia2009lte, yu2007transmitter, boccardi2006zero, vu2011miso}.

We consider IA in two types of DAS, ones in which individual RRU powers are upper bounded (called \emph{maximum power constraints}) and others in which all RRUs must transmit at a constant power level (called \emph{strict power constraints}) \cite{yu2007transmitter}. This paper thus completes the preliminary results presented in \cite{starr2011interference} which focused on the special case of IA with per-antenna power constraints in co-located antenna systems. To satisfy maximum power constraints, we consider a simple strategy of transmit power back-off and show that the back-off procedure incurs a systematic loss in sum-rate. To gain a better quantitative understanding of the sum-rate loss due to back-off, we give an analytical characterization of the back-off factor's statistics under the simplifying assumption of channels with equal pathloss, i.e., in the case of co-located antennas. The development of more sophisticated IA strategies that satisfy maximum power constraints and avoid the loss incurred by back-off is an interesting topic for future work. In systems where RRUs must transmit at a fixed power level, we show that the addition of such strict power constraints negatively affects the feasibility of IA. In other words, realizing IA with strict power constraints requires more antennas at the transmitters or receivers. To examine this reduction in feasibility, we leverage the methodology in \cite{yetis2010feasibility} to derive properness conditions for IA systems with strict per-RRU power constraints. While properness and feasibility are in general not equivalent, the results of \cite{gonzalez2012feasibility, razaviyayn2011degrees, bresler2011settling} indicate that proper systems are most often feasible except in a few corner cases. As a result, properness can provide a simple and sufficiently accurate predictor of IA feasibility. To further demonstrate the true feasibility of IA in proper systems, we present a simple iterative IA algorithm based on the alternating minimization solution in \cite{peters2011cooperative, gomadam2008approaching}. While more sophisticated algorithms are an interesting topic for future work, the proposed solution is shown to work well in simulation. Namely, the proposed algorithm avoids the sum-rate loss of power back-off and interestingly achieves the same performance as unconstrained IA. Finally, simulations that incorporate realistic channel conditions by using the 3rd Generation Partnership Program's Spatial Channel Model \cite{salo2005matlab,3gpp2003scm} indicate that the proposed DAS with IA concept constitutes a promising system-level solution which outperforms two benchmarks: (i) IA in traditional cellular systems, and (ii) DAS with existing transmission strategies.

This paper is organized as follows. Section \ref{sec:system_model} introduces the system model and Section \ref{sec:ia_summary} provides the relevant background on IA. Section \ref{sec:MaxConstraints} investigates the performance of IA with maximum per-RRU power constraints. Section \ref{sec:StrictConstraints} investigates the feasibility of, and provides a candidate algorithm for, implementing IA with strict per-RRU power constraints. Section \ref{sec:Simulations} provides corroborative numerical results on IA with per-RRU power constraints and investigates its performance in a realistic cellular system model with distributed antennas. Finally, Section \ref{sec:Conclusion} concludes the paper. 

We use the following notation throughout this paper: $\bA$ is a matrix, $\ba$ is a vector, $a$ is a scalar, and $\mathcal{A}$ is a set. The function $\mathrm{rank}(\bA)$ denotes the rank of $\bA$, $\|\bA\|_F$ is the Frobenius norm of $\bA$, and $\|\ba\|_2$ is the two-norm of vector $\ba$. $\bA(m, n)$ is the element of $\bA$ in the $m^\mathrm{th}$ row and $n^\mathrm{th}$ column and $\left[\bA\right]_{1:N,:}$ ($\left[\bA\right]_{:,1:N}$) are the first $N$ rows (columns) of $\bA$. $s_\mathrm{min} (\bA)$ and $s_\mathrm{max} (\bA)$ are the minimum and maximum singular values of $\bA$ respectively. $\bI_N$ is the $N \times N$ identity matrix, $\mathbf{1}_N$ is the $N$-length column vector composed of all ones, and $\mathcal{CN} (\bm,\bR)$ is the complex Gaussian distribution of mean $\bm$ and variance $\bR$. The notation $a \stackrel{(d)}{=} b$ means that $a$ and $b$ are equivalent in distribution. Finally, $\mathcal{Q}(x; r)$ denotes the cumulative distribution function (CDF) of the chi-squared distribution with $r$ degrees of freedom evaluated at $x$.

\section{System Model} \label{sec:system_model}

\begin{figure}[t!]
	\centering
    \includegraphics[width=6in]{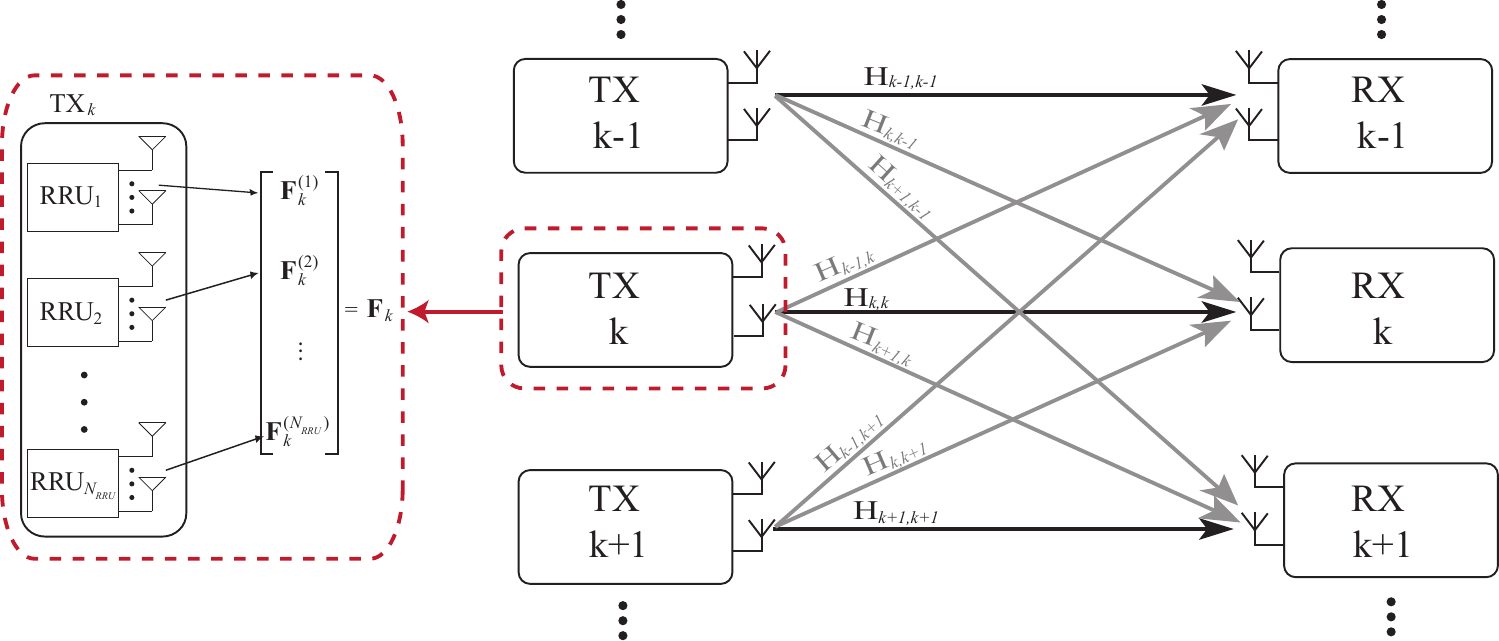}
	\caption{$K$-User MIMO interference channel model with distributed antenna transmitters. Each transmitter consists of $\Nrru$ remote radio units each with $\Nt/\Nrru$ antennas.}
  \label{fig:InterferenceChannel}
\end{figure}

Consider the $K$-user interference channel shown in Fig. \ref{fig:InterferenceChannel} in which transmitter $k$ communicates with its targeted receiver $k$ and interferes with all other receivers $\ell \neq k$. We assume that the system is symmetric, meaning that the number of transmit antennas $\Nt$, receive antennas $\Nr$, and data streams $\Ns$ is the same for all transmitter-receiver pairs. We denote such symmetric systems as $(\Nt \times \Nr,\ \Ns)^K$. In our DAS setup, we assume that each transmitter's
antennas are distributed among $\Nrru$ remote radios, each with $\Nt/\Nrru$ antennas as shown in Fig. \ref{fig:InterferenceChannel}.

Let us define $\bx_k$ to be the $\Ns \times 1$ symbol vector transmitted by node $k$ with $\mathbb{E}\left[\bx_k\bx_k^*\right] = \bI_\Ns$, $\bF_k$ to be the $\Nt\times\Ns$ precoding matrix used by transmitter $k$, $\bH_{k\ell}$ to be the $\Nr\times\Nt$ channel from transmitter $\ell$ to receiver $k$, and $\bz_k$ to be $\Nr\times 1$ vector of i.i.d complex Gaussian noise observed at receiver $k$ with covariance matrix $\sigma^2 \bI_\Nr$. Assuming perfect time and frequency synchronization, user $k$'s received signal, $\by_k$, can be written as
\begin{equation}
\by_k=\bH_{kk}\bF_{k}+\sum\limits_{\ell \neq k}\bH_{k\ell}\bF_{\ell}+\bz_k.
\label{eqn:ReceivedSignal}
\end{equation}
The statistical properties of the channels $\bH_{k\ell}$ are given in detail later, as needed by the different results in Sections \ref{sec:MaxConstraints}, \ref{sec:StrictConstraints} and \ref{sec:Simulations}. To enable the calculation of the IA precoders, however, the channels are assumed to be known perfectly to both transmitters and receivers. In practice, channel knowledge can be obtained by reciprocity or feedback \cite{Ayach2010, thukral2009interference}. 

The precoders $\bF_k$ are constrained to satisfy a total power constraint given by $\|\bF_k\|_F^2= P$, where $P$ is the \emph{total transmit power} available at each transmitter. Total available power, however, is not the only constraint on the transmit precoders $\bF_k$. Practical power amplifiers and transceiver architectures typically place a limit on the power radiated by \emph{individual antennas} or by \emph{groups of antennas} located at the same RRU. To handle per-RRU power constraints mathematically, we partition the precoders $\bF_k$ into $\Nrru$ transmit subfilters $\bF^{(r)}_k$ each of size $\frac{\Nt}{\Nrru}\times \Ns$ such that $\bF_k = \left[\bF^{(1)*}_k, \bF^{(2)*}_k, \hdots ,\bF^{(\Nrru)*}_k \right]^*$ as illustrated in Figure \ref{fig:InterferenceChannel}. Constraints are then placed on the Frobenius norm of each subfilter $\bF^{(r)}_k$. In cases where \emph{per-antenna} power is restricted, constraints are simply placed on the norm of individual rows $\bff^{(r)}_k$ of the precoder $\bF_k$ (equivalently RRUs can be thought of as having a single antenna each). With this notation, we define the two power constraints:
\begin{enumerate}
\item \emph{Maximum Power Constraints} where the maximum power radiated by an RRU or an antenna is constrained, i.e., $\|\bF^{(r)}_k\|^2_F \leq \frac{P}{\Nrru}$ in the case of per-RRU constraints and $\|\bff^{(r)}_k\|_2^2 \leq \frac{P}{\Nt}$ in the case of per-antenna constraints.
\item \emph{Strict Power Constraints} where a strict equality constraint is placed on the power radiated by different RRUs or antennas, i.e., $\|\bF^{(r)}_k\|^2_F = \frac{P}{\Nrru}$ in the case of per-RRU constraints or  $\|\bff^{(r)}_k\|_2^2 = \frac{P}{\Nt}$ in the case of per-antenna constraints.
\end{enumerate}
Sections \ref{sec:MaxConstraints} and \ref{sec:StrictConstraints} analyze the effects of these more stringent constraints on IA feasibility and performance. When IA's performance is considered, the main metric of interest is the average sum-rate achieved with complex Gaussian signaling and treating interference as noise. Under these assumptions, sum-rate is given by \cite{blum2003mimo}
\begin{equation}
\Rsum=\mathbb{E}_\mathcal{H} \left[\sum\limits_{k=1}^{K}\log_2\left|\bI_\Nr+\left(\sigma^2\bI_\Nr+\sum\limits_{\ell \neq k}\bH_{k\ell}\bF_{\ell}\bF_{\ell}^*\bH_{k\ell}^*\right)^{-1}\bH_{kk}\bF_k\bF_k^*\bH_{kk}^*\right|\right],
\end{equation}
where $\mathbb{E}_\mathcal{H}$ denotes expectation over the distributions of the channels $\bH_{k\ell}\ \forall k,\ell$.

\section{Interference Alignment} \label{sec:ia_summary}

The objective of IA is to limit the dimension of the interference observed at each receiver by packing multiple interfering signals into a reduced dimensional subspace. This ensures that there exists a receive subspace, of adequate dimension, in which desired signals can be observed interference-free. It is this dimensionality reduction that allows IA to achieve the degrees of freedom of some frequency-selective interference channels~\cite{cadambe2008interference, gou2010degrees}, and achieve good sum-rate performance in constant MIMO channels~\cite{peters2011cooperative}. 

While IA can be used with any receiver architecture, alignment can be intuitively understood by examining the operation of a linear interference zero-forcing receiver. Define $\bW_k$ to be the $\Nr \times \Ns$ linear zero-forcing combiner used by receiver $k$. The received signal at the output of $\bW_k$ is given by 
\begin{equation} \label{eqn:zero_forcing}
  \mathbf{W}_{k}^{*}\mathbf{y}_{k} = \mathbf{W}_{k}^{*} \mathbf{H}_{kk} \mathbf{F}_{k} \mathbf{x}_{k} + \mathbf{W}_{k}^{*} \sum_{\ell \ne k}\mathbf{H}_{k\ell} \mathbf{F}_{\ell} \mathbf{x}_{\ell} + 
    \mathbf{W}_{k}^{*} \mathbf{z}_{k}.
\end{equation}
At the output of $\bW_k$, the conditions that ensure perfect alignment can be stated as
\begin{align}
  \mathbf{W}_{k}^{*} \mathbf{H}_{k\ell} \mathbf{F}_{\ell} &= \mathbf{0}_{\Ns}, & \forall k, \ell \in \mathcal{K},\ k \neq \ell, 
  \label{eqn:ia_formulation:002} \\
  \mathrm{rank}\left(\mathbf{W}_{k}^{*} \mathbf{H}_{kk} \mathbf{F}_{k}\right) &= N_{s}, & \forall k \in \mathcal{K}, 
  \label{eqn:ia_formulation:001}
\end{align}
where $\mathcal{K}$ is the set of users and $\mathbf{0}_{\Ns}$ is the $\Ns \times \Ns$ all-zeros matrix. Interference alignment and cancellation is guaranteed by (\ref{eqn:ia_formulation:002}), and (\ref{eqn:ia_formulation:001}) ensures the decodability of user $k$'s desired signal. The IA conditions in (\ref{eqn:ia_formulation:002})-(\ref{eqn:ia_formulation:001}) have been used extensively in the literature to derive several IA algorithms~\cite{gomadam2008approaching, peters2011cooperative, santamaria2010maximum} and various performance results~\cite{yetis2010feasibility, razaviyayn2011degrees, thukral2009interference, Ayach2010, nosrat2010mimo}.

The work in \cite{yetis2010feasibility} leveraged conditions (\ref{eqn:ia_formulation:002}) and (\ref{eqn:ia_formulation:001}) to characterize the systems in which IA is feasible. The authors of \cite{yetis2010feasibility} defined the notion of \emph{proper IA systems} in which the number of free variables in the transmit precoders $\bF_k$ always exceeded the number of constraints imposed by condition (\ref{eqn:ia_formulation:002}). It was then shown that system properness constitutes a necessary condition for IA feasibility. More recent work has shown that, while properness does not rigorously guarantee IA feasibility, proper systems are most often feasible except in a few corner cases~\cite{gonzalez2012feasibility, razaviyayn2011degrees}. As a result, the notion of properness can provide a sufficiently accurate predictor of IA feasibility. For the symmetric systems considered in this paper, for example, an IA system is considered proper (and most likely feasible) as long as
\begin{equation} \label{eqn:ia_feasibility_conditions}
\Nt +\Nr\geq \left( K + 1 \right)\Ns.
\end{equation}
The conditions in (\ref{eqn:ia_formulation:002})-(\ref{eqn:ia_formulation:001}) have similarly been used to derive iterative algorithms in cases where closed form solutions do not exist. In the alternating minimization solution of \cite{peters2011cooperative}, the total power of \emph{leakage interference}, defined as  $\sum_\ell \|\bW_k\bH_{k\ell}\bF_\ell\|^2$, is iteratively minimized over alternating choices of $\mathbf{F}_{\ell}$ and $\mathbf{W}_{k}$. Using the derived algorithms, IA was ultimately shown to provide good high-SNR sum-rate in a variety of MIMO interference channels. 

Since existing IA algorithms, feasibility results, and performance analysis do not consider per-RRU or per-antenna power constraints, it remains unclear if IA's promise carries over to DAS. To see this, note that algorithms that neglect per-RRU power constraints yield precoders with an unbalanced power profile across different antennas. This power imbalance can be significant in DAS where different antennas experience significantly different pathloss. If maximum power is constrained, transmitters can back-off their \emph{total transmit power} to ensure that no RRU exceeds its power constraint. Power back-off, however, will result in a loss of both effective SNR and IA sum-rate which we characterize analytically in Section \ref{sec:MaxConstraints}. In systems that require all RRUs to transmit at the \emph{same power}, existing IA algorithms cannot be used altogether and power back-off cannot be used to balance RRU transmit power. In fact, Section \ref{sec:StrictConstraints} shows that placing such strict constraints on per-RRU power significantly affects IA feasibility. For such systems, Section \ref{sec:StrictConstraints} derives revised properness conditions and provides an iterative algorithm that enables calculating constrained IA precoders.

\section{Interference Alignment with Maximum Power Constraints} \label{sec:MaxConstraints}

When a limit is placed on per-RRU transmit power, IA precoders must satisfy the following $\Nrru$ inequality constraints 
\begin{equation} 
  \| \bF_k^{(r)}\|^2_F \leq \frac{P}{\Nrru},\qquad \forall r \in  \rruset, \ \forall k \in \mathcal{K},
  \label{eqn:MaxConstraintsFormulation}
\end{equation} 
where $\mathcal{R}$ is the set of RRUs. Since per-antenna constraints are mathematically equivalent to (\ref{eqn:MaxConstraintsFormulation}) with $\Nrru = \Nt$, we focus on the general case of per-RRU constraints. Examining the IA conditions given by \eqref{eqn:ia_formulation:002} and \eqref{eqn:ia_formulation:001}, we note that the constraints in (\ref{eqn:MaxConstraintsFormulation}) should not affect IA feasibility. Indeed, suppose that there exists a set of precoders $\bF_k$ and combiners $\bW_k$ that satisfy (\ref{eqn:ia_formulation:002})-(\ref{eqn:ia_formulation:001}) with a maximum RRU transmit power of $\beta_{k} = \max_{r \in \mathcal{R}} \|\bF_{k}^{(r)}\|_F$. In this case, the \emph{scaled} precoders $\frac{\sqrt{P/\Nrru}}{\beta_{k}} \bF_k$ simultaneously ensure that alignment is preserved and transmit power constraints are respected.

While the constraints in (\ref{eqn:MaxConstraintsFormulation}) do not affect feasibility, satisfying them via power back-off causes a systematic reduction in IA sum-rate since $\frac{\sqrt{P/\Nrru}}{\beta_k} < 1$ with probability one\footnote{This can be easily seen by noting that $\frac{\sqrt{P/\Nrru}}{\beta_{k}}\leq 1$ by definition and equality is achieved if the IA precoders $\bF_k$ are such that the continuous random variables $\|\bF^{(r)}_k\|^2_F$ are such that $\|\bF^{(r)}_k\|^2_F = \frac{P}{\Nrru} \forall r = 1, \hdots, \Nrru$. Equality happens with probability 0 for any existing IA algorithm which does not explicitly attempt to balance RRU transmit powers.}. To gain a quantitative understanding of the performance degradation induced by power back-off, we examine the mean loss in sum-rate at high SNR which we define in the following proposition.
\begin{proposition} \label{prop:RateLoss}
The mean loss in sum-rate resulting from transmit power back-off at high SNR is given by 
\begin{equation}
\bar{R}_\mathrm{loss} = \sum\limits_{k=1}^{K} \Ns \mathbb{E}\left[ \log_{2}\left( \frac{P}{\Nrru \beta_k^2}\right) \right].
\label{eqn:RateLoss}
\end{equation}
\end{proposition}
\begin{IEEEproof}
A brief derivation is provided in Appendix \ref{append:inequality_sum_rate_loss}.
\end{IEEEproof}

Given Lemma \ref{prop:RateLoss}, all that remains to complete the characterization of $\bar{R}_\mathrm{loss}$ is to derive the statistics of the random variables $\beta_k^2$. The distribution of $\beta^2_k$, however, is tied to the distribution of the precoders $\bF_k$ and is thus dependent (non-trivially) on the statistical model used for the propagation channels $\bH_{k\ell}$. This makes deriving the distribution of $\beta^2_k$ for general channel models intractable. Thus, to simplify the rate analysis, we make the following channel assumption.
\begin{assumption}
We assume that all channels $\bH_{k\ell}$ are Rayleigh fading, i.e., have i.i.d $\mathcal{CN}(0,1)$ entries.
\label{as:rayleigh}
\end{assumption}
While Assumption \ref{as:rayleigh} neglects the large scale fading present in DAS, it helps in giving an example of the non-negligible effect of power back-off in systems with co-located antennas, i.e., in systems that strictly follow Assumption \ref{as:rayleigh}. Mathematically, Assumption \ref{as:rayleigh} enables us to use the following result from \cite{nosrat2010mimo}.
\begin{theorem} \label{thm:rayleigh_precoder_distribution}
Assuming Rayleigh-fading channels, the precoders $\bF_k \in \mathbb{C}^{\Nt\times \Ns}$ for $k \in \mathcal{K}$ generated using the IA algorithms in \cite{cadambe2008interference, peters2011cooperative, gomadam2008approaching} are Haar-distributed, i.e., they are uniformly distributed over the set of orthogonal $\Ns$-frames in $\mathbb{C}^{\Nt}$.
\end{theorem}
\begin{IEEEproof}
Given in~\cite{nosrat2010mimo}.
\end{IEEEproof}
Theorem \ref{thm:rayleigh_precoder_distribution} facilitates the derivation of $\beta_k$'s statistics by (i) identifying a single tractable distribution for the precoders $\bF_k$, and (ii) consequently indicating that all variables $\beta_k$ are statistically equivalent. As a result, we henceforth drop the subscript $(\cdot)_k$ from $\beta_k$.

For simplicity of exposition, we start by analyzing the case of $\Ns = 1$ and then apply results from random matrix theory to generalize our analysis to the case of $\Ns > 1$. For the case of $\Ns = 1$, and as a result of Theorem \ref{thm:rayleigh_precoder_distribution}, we leverage the following proposition on Haar distributed random vectors.
\begin{proposition}
Let $\bff \in \mathbb{C}^{\Nt\times 1}$ be a random beamforming vector generated from a Haar (uniform) distribution over the $\Nt$-dimensional hypersphere and let $\mathcal{Q}(x; r)$ denote the CDF of the chi-squared distribution with $r$ degrees of freedom. Further, let $\bff^{(r)}\ \forall r \in \mathcal{R}$ denote the sub-vector corresponding to the beamforming weights used by the antennas in the $r^\mathrm{th}$ RRU. Then, if $\frac{\Nt}{\Nrru}$ is a constant natural number, the CDF of the random variable $\beta^2 = \mathrm{max}_{r\in \mathcal{R}} \|\bff^{(r)}\|_2^2$ satisfies the following limit
\begin{equation}
\lim_{\Nt,\Nrru \to \infty} \frac{1}{\Nt}\log\mathbb{P}_{\beta^2}\left\{\beta^2\leq \frac{x}{\Nt}\right\}=\log \mathcal{Q}\left(x; \frac{\Nt}{\Nrru}\right),
\end{equation}
which implies that, for large $\Nt$, the CDF of $\beta^2$ can be approximated as
\begin{equation}
\mathbb{P}_{\beta^2}\left\{\beta^2\leq x\right\}\approx \mathcal{Q}\left(\Nt x; \frac{\Nt}{\Nrru}\right)^{\Nt}.
\label{eqn:CDFApproximation}
\end{equation}
\label{prop:CDFApproximation}
\end{proposition}
\begin{IEEEproof}
  Given in Appendix \ref{append:inequality_max_row_norm_cdf_beamforming}.
\end{IEEEproof}

Proposition \ref{prop:CDFApproximation} thus gives a complete characterization of the power back-off factor affecting single-stream IA systems with maximum power constraints. To generalize Proposition \ref{prop:CDFApproximation} to the case of $\Ns > 1$, we leverage the following result from \cite{vershynin2010introduction} which states that for large $\Nt$, the $\Nt \times \Ns$ Haar distribution converges to the $\Nt \times \Ns$ Gaussian distribution.
\begin{proposition}
Let $\bF$ be a Haar distributed $\Nt\times\Ns$ unitary matrix and $\bX$ be a random $\Nt\times\Ns$ Gaussian matrix. Then for large $\Nt$ we have that $\frac{1}{\sqrt{\Nt}}\bX\stackrel{(d)}{=}\bF$. Stated otherwise, the process of calculating $\Nt \times \Ns$ unitary IA precoders yields matrices that are statistically equivalent to generating $\Nt \times \Ns$ Gaussian matrices and scaling them by the constant factor $1/\sqrt{\Nt}$.
\label{prop:CDFApproximation2}
\end{proposition}
\begin{IEEEproof}
The proof follows from the limiting behavior of tall Gaussian matrices~\cite{tulino2004random} and can be found in~\cite{vershynin2010introduction}.
\end{IEEEproof}
From Proposition \ref{prop:CDFApproximation2}, we conclude that the distribution of the maximum RRU transmit power $\beta^2 = \mathrm{max}_{r\in\mathcal{R}}\|\bF^{(r)}\|_F^2$ can be derived by instead examining the random variable $\gamma = \mathrm{max}_{r \in \mathcal{R}} \|\bX^{(r)}\|_F^2$ where $\bX^{(r)}$ is a sub-matrix of the Gaussian matrix $\bX$. Since $\bX$ is an $\Nt\times\Ns$ Gaussian matrix, the diagonal elements of $\bX\bX^*$ are chi-squared distributed with $\Ns$ degrees of freedom. Therefore, the CDF of $\beta^2$ can be approximated as
\begin{equation}
\mathbb{P}_{\beta^2}\left\{\beta^2 \leq x \right\} \approx \mathcal{Q}\left(\Nt x ; \frac{\Ns\Nt}{\Nrru}\right)^{\Nt}.
\label{eqn:CDFApproximation2}
\end{equation}
Having found the distribution of the power back-off factor $\beta^2$ for all $\Ns \geq 1$, evaluating the mean loss in sum-rate give in Proposition \ref{prop:RateLoss} reduces to a simple one-dimensional integral that can be evaluated numerically. In Section \ref{sec:Simulations} we provide numerical results demonstrating the accuracy of the derived back-off statistics in predicting IA performance in the presence of maximum power constraints.

We note that it may be possible to develop more sophisticated methods or algorithms to achieve alignment with maximum power constraints. In particular, it may be possible to develop algorithms that overcome the sum-rate loss incurred by the back-off strategy considered in this section. The derivation of such strategies is left for future work.

\section{Interference Alignment with Strict Power Constraints} \label{sec:StrictConstraints}

In the previous section, we argued that while IA with maximum power constraints is always feasible, transmit power back-off results in a systematic loss of sum rate. Thus, in practical systems that operate in low or medium SNR environments, it may be advantageous to design algorithms that avoid power back-off and ensure that all system resources are fully exploited. 

To ensure that systems transmit at full power, strict equality constraints can be placed on the IA precoders $\bF_k$. Recalling the definition in Section \ref{sec:system_model}, such constraints can be written as
\begin{equation}
\|\bF_k^{(r)}\|_F^2=\mathrm{trace}\left(\bF_k^{(r)}\bF_k^{(r)*}\right)=\frac{P}{\Nrru},\ \forall r\in \mathcal{R}, \forall k \in \mathcal{K}.
\label{eqn:StrictConstraintsFormulation}
\end{equation}
As stated in Section \ref{sec:MaxConstraints}, if per-RRU transmit power is neglected, IA algorithms will generate precoders in which the quantities $\|\bF^{(r)}_k\|_F$ are continuous random variables. Thus for existing IA solutions $\|\bF^{(r)}_k\|^2_F \neq \frac{P}{\Nrru}$ with probability one. Unlike maximum power constraints, however, (\ref{eqn:StrictConstraintsFormulation}) cannot be satisfied via simple transmit power back-off. Further, it is possible that enforcing the strict power constraints in (\ref{eqn:StrictConstraintsFormulation}) will fundamentally affect the feasibility of IA in DAS and will necessitate the development of improved IA algorithms that explicitly account for per-RRU power. The remainder of this section is devoted to exploring both feasibility and algorithm design in the presence of strict power constraints.

\subsection{The Feasibility of IA with Strict Power Constraints} \label{sec:IAFeasibility}

The problem of determining the feasibility of IA over constant MIMO channels has been
studied extensively in \cite{razaviyayn2010linear, razaviyayn2011degrees, yetis2010feasibility, gonzalez2012feasibility, bresler2011settling}. Characterizing feasibility was shown to require a combination of algebraic geometry and differential topology. The derived feasibility results are in the form of algorithmic tests of polynomial complexity that cannot be reduced to simple requirements on system parameters such as $K$, $\Ns$, $\Nt$, and $\Nr$. The results of \cite{gonzalez2012feasibility}, however, indicate that the simpler notion of \emph{system properness} is often a sufficiently accurate predictor of IA feasibility, even though properness and feasibility are not rigorously connected in general~\cite{yetis2010feasibility, razaviyayn2011degrees, bresler2011settling}. System properness was first introduced in \cite{yetis2010feasibility} to conjecture that IA is feasible if the underlying system of equations in (\ref{eqn:ia_formulation:002}) is proper\footnote{The equations in (\ref{eqn:ia_formulation:001}) need not be considered as they are automatically satisfied with high probability for continuous channel distributions of interest.}. A system is called ``proper'', if for any subset of equations from (\ref{eqn:ia_formulation:002}), the number of unsatisfied scalar equations \emph{does not exceed} the number of free variables involved in that subset. As a result, the concept of properness can provide ``feasibility'' conditions based solely on arithmetic relationships between $K$, $\Ns$, $\Nt$, and $\Nr$. Thus, properness can serve as a simple and accurate proxy for general feasibility tests.


In the remainder of this section, we extend the notion of properness to
systems with strict per-RRU or per-antenna power constraints. Since the MIMO interference channels considered in this paper are symmetric, properness can be determined by examining the number of free variables and unsatisfied equations in the \emph{full set of IA conditions} (\ref{eqn:ia_formulation:002}), i.e., there is no need to consider subsets of equations from (\ref{eqn:ia_formulation:002}) \cite{yetis2010feasibility}. Let $N_\mathrm{v}$ be the total number of free variables in the IA system, and let $N^{(1)}_\mathrm{e}$ and $N^{(2)}_\mathrm{e}$ be the number of non-trivially satisfied equations in (\ref{eqn:ia_formulation:002}) and (\ref{eqn:StrictConstraintsFormulation}) respectively. 

From \cite{yetis2010feasibility}, the number of free variables $N_\mathrm{v}$ and unsatisfied equations $N^{(1)}_\mathrm{e}$ in (\ref{eqn:ia_formulation:002}) are given by
\begin{align}
N_\mathrm{v} & = K (\Nt + \Nr - 2\Ns)\Ns, \label{eqn:Nv}\\
N^{(1)}_\mathrm{e} & = K (K - 1)\Ns^2. \label{eqn:Ne1}
\end{align}
The details of this counting argument are provided in \cite{yetis2010feasibility}. In short, $N_\mathrm{v}$ is determined by
counting the free variables in $\bF_k\ \forall k\in \mathcal{K}$ and $\bW_k\ \forall k\in \mathcal{K}$ after reducing them to their unique basis representations, while $N^{(1)}_\mathrm{e}$ is determined by counting the scalar equations in (\ref{eqn:ia_formulation:002}). To count the number of non-trivially satisfied equations in (\ref{eqn:StrictConstraintsFormulation}), we provide the following result. 

\begin{proposition} 
The number of non-trivially satisfied equations in (\ref{eqn:StrictConstraintsFormulation}) is given by 
\begin{equation}
N^{(2)}_e = \mathrm{max}\left\{K(\Nrru - \Ns) , 0\right\} . 
\label{eqn:Ne2}
\end{equation}
\label{prop:NumConstraints}
\end{proposition}
\begin{IEEEproof}
Given in Appendix \ref{append:papc_equality_properness}.
\end{IEEEproof}
Combining (\ref{eqn:Nv}), (\ref{eqn:Ne1}) and (\ref{eqn:Ne2}), the following characterization of system properness for IA with strict per-RRU power constraints can be obtained.
\begin{proposition}
A symmetric IA system with strict per-RRU power constraints is proper, and is thus expected to be feasible, 
if and only if
\begin{equation}
(\Nr + \Nt)\Ns \geq (K + 1)\Ns^2 +\mathrm{max}\left\{(\Nrru - \Ns),0\right\}
\label{eqn:IAMaxConstraintsProperness}
\end{equation}
\label{prop:IAMaxConstraintsProperness}
\end{proposition}
\begin{IEEEproof}
The result follows immediately by substituting (\ref{eqn:Nv}), (\ref{eqn:Ne1}) and (\ref{eqn:Ne2}) into the definition of system properness, i.e., $N_\mathrm{v} \geq N^{(1)}_\mathrm{e} + N^{(2)}_\mathrm{e}$.
\end{IEEEproof}

\begin{figure}[t!]
	\centering
    \includegraphics[width=6in]{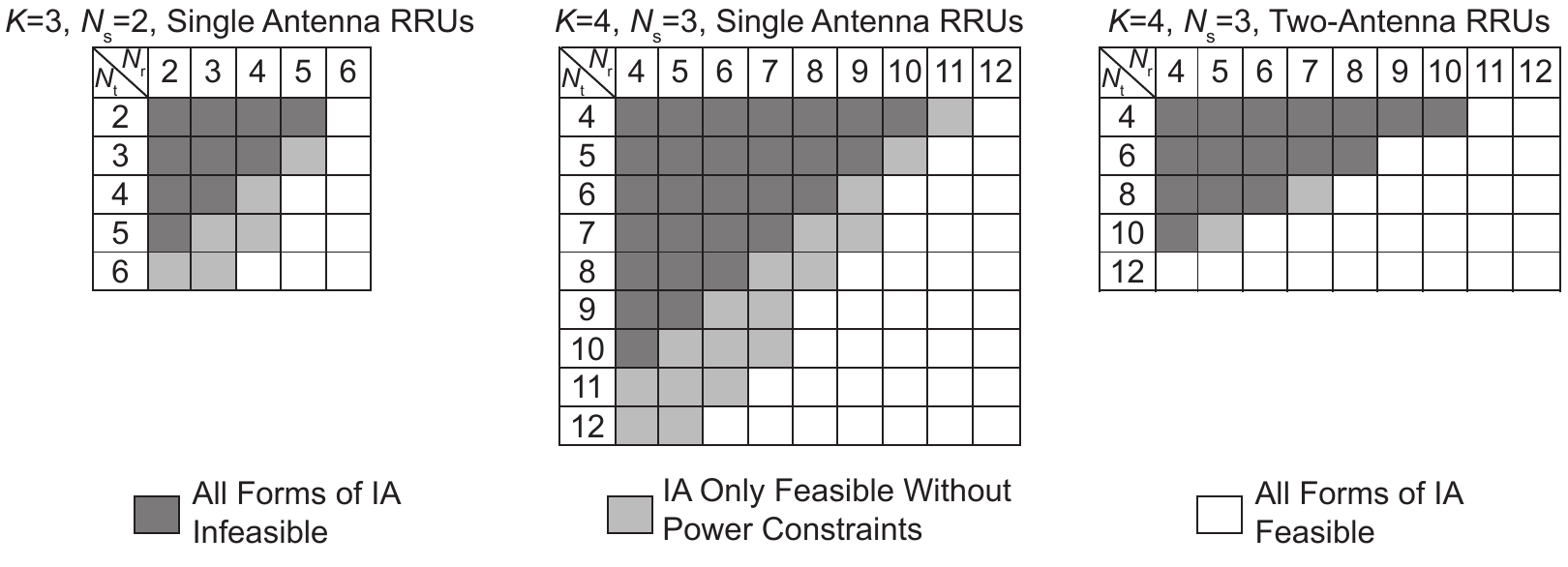}
	\caption{Tables highlighting network configurations in which IA is “infeasible”, strictly “feasible”, or “feasible” only without per-RRU or per-antenna power constraints as predicted by the derived conditions on system properness.}
  \label{fig:FeasibilityTables}
\end{figure}

Examining the properness conditions derived in Proposition \ref{prop:IAMaxConstraintsProperness} for IA with strict per-RRU power constraints, we make the following observations:
\begin{enumerate}
\item For single antenna RRU's, the properness condition in (\ref{eqn:IAMaxConstraintsProperness}), simplifies to the per-antenna constrained case examined in \cite{starr2011interference}. Namely, since $\Nrru = \Nt \geq \Ns$, the condition for properness simplifies to
\begin{equation}
(\Nr + \Nt)\Ns \geq (K + 1)\Ns^2 + (\Nt - \Ns).
\end{equation}
\item Interestingly, per-RRU power constraints only reduce feasibility in cases where $\Nrru > \Ns$. In cases where $\Nrru \leq \Ns$, the properness condition in (\ref{eqn:IAMaxConstraintsProperness}) reduces to the traditional IA properness condition in (\ref{eqn:ia_feasibility_conditions}). As a special case, this regime includes traditional co-located antenna systems ($\Nrru = 1$) with only a total transmit power constraint.
\item Per-RRU power constraints destroy the symmetry of alignment since the properness condition no longer depends on $\Nt$ and $\Nr$ only through their sum $\Nt+\Nr$. Therefore, unlike in the traditional IA case explored in \cite{yetis2010feasibility}, if an $(\Nt \times \Nr,\Ns)^K$ system is proper and thus likely feasible, the  $\left((\Nt+1)\times(\Nr-1),\Ns\right)^K$ system need not be feasible. The reciprocal $(\Nr\times\Nt,\Ns)^K$ system need not be feasible either.
\item Interestingly, in the case of $\Ns = 1$ with single antenna RRUs, i.e., $\Nrru = \Nt$, transmit antennas are entirely useless for alignment or any transmit-side interference nulling. This can be seen by noting that, in this case, the properness condition simplifies to $(\Nt + \Nr) \geq (K + 1)+(\Nt - 1) \Longrightarrow \Nr \geq K$. The condition $ \Nr \geq K$ implies that the receivers must have enough antennas to “single-handedly” cancel all unaligned interference and decode their desired signals. Note that in this case, the transmit precoders are simply equal-gain beamforming vectors \cite{love2003equal} which we have now shown can never be used for alignment.
\item The condition in (\ref{eqn:IAMaxConstraintsProperness}) confirms the intuition that having multiple antenna RRUs (such that $\Nrru < \Nt$) significantly reduces the effect of per-RRU power constraints and thus improves IA feasibility.
\end{enumerate}

To get a better understanding of the effect of per-RRU power constraints on IA feasibility, Fig. \ref{fig:FeasibilityTables} tabulates some example scenarios in which IA is always feasible, feasible \emph{only in the absence of per-RRU power constraints}, or always infeasible,. Comparing the two $K = 4$ cases with single and multi-antenna RRUs, we see that for a fixed number of transmit antennas $\Nt$, even as little as two antennas per-RRU can dramatically improve feasibility.

\subsection{Algorithm for IA with Strict per-RRU Power Constraints} \label{sec:IAAlgorithm}

While Proposition \ref{prop:IAMaxConstraintsProperness} gives properness conditions under which IA with strict per-RRU power constraints is expected to be possible, it provides no insight on how to realize such alignment precoders. To better demonstrate the feasibility of IA with per-RRU constraints, we extend the method of alternating minimization used in \cite{peters2011cooperative, gomadam2008approaching} to provide a simple algorithm which we show in Section \ref{sec:Simulations} satisfies both the alignment conditions in (\ref{eqn:ia_formulation:002}) and the per-RRU constraints in (\ref{eqn:StrictConstraintsFormulation}).

The alternating minimization strategy used in \cite{peters2011cooperative, gomadam2008approaching} can be summarized as iteratively minimizing the total power of leakage interference, defined as 
\begin{equation}
\mathcal{J}_\mathrm{IL}\left(\left\{\bF_k\right\}_{k\in \mathcal{K}},\left\{\bW_k\right\}_{k\in\mathcal{K}}\right) = \sum_{\ell \in \mathcal{K}}\sum_{k\in \mathcal{K}\backslash \ell} \|\bW_k^*\bH_{k\ell}\bF_{\ell}\|_F^2,
\end{equation}
over alternating choices of $\bW_k$ and $\bF_k$. We refer the reader to \cite{peters2011cooperative, gomadam2008approaching} for a detailed derivation of the optimal choice of $\bW_k$ and $\bF_k$ in each iteration and give the final result here for brevity. At each iteration, the combiners $\bW_k$ are chosen as
\begin{equation}
\bW_k = \nu_\mathrm{min}^\Ns \left(\sum\limits_{\ell \neq k} \bH_{k\ell}\bF_\ell\bF_\ell^*\bH_{k\ell}^*\right),
\label{eqn:CombinerUpdate}
\end{equation}
followed by an update to the choice of precoders given by
\begin{equation}
\bF_k = \nu_\mathrm{min}^\Ns \left(\sum\limits_{\ell \neq k} \bH_{\ell k}^*\bW_\ell\bW_\ell^*\bH_{\ell k}\right),
\label{eqn:PrecoderUpdate}
\end{equation}
where $\nu_\mathrm{min}^\Ns (\cdot)$ denotes the $\Ns$ least dominant eigenvectors of a symmetric matrix. To satisfy the per-RRU power constraints, we add a projection step onto the set of precoders with a fixed RRU transmit power of $\frac{P}{\Nrru}$. The projection on that set amounts to scaling and updating the blocks of $\bF_k$ as 
\begin{equation}
\bF^{(r)}_k \leftarrow \frac{\sqrt{\frac{P}{\Nrru}}}{\|\bF_k^{(r)}\|_F}\bF^{(r)}_k.
\end{equation}
The steps of the algorithm are listed more formally in Algorithm \ref{alg:generate_papc_equality_precoders}.

\begin{algorithm}[t!] \label{alg:generate_papc_equality_precoders}
  \caption{IA with Strict per-RRU Power Constraints}
  \begin{algorithmic}
    \REQUIRE $\left\{ \mathbf{H}_{k\ell} \right\}_{k, \ell =1}^{K}$
    \ENSURE  $\left\{ \mathbf{F}_{k} \right\}_{k=1}^K$ and $\left\{ \mathbf{W}_{k} \right\}_{k=1}^K$
    \STATE Arbitrarily generate initial $\left\{ \mathbf{F}_{k} \right\}_{k=1}^K$ and $\left\{ \mathbf{W}_{k} \right\}_{k=1}^K$
    \FOR{a fixed number of iterations}
      \STATE Construct $\left\{\bW_k\right\}_{k=1}^K$ using (\ref{eqn:CombinerUpdate}).
      \STATE Construct $\left\{\bF_k\right\}_{k=1}^K$ using (\ref{eqn:PrecoderUpdate}).      
      \FOR{$\forall k \in \mathcal{K}$ and $r \in \mathcal{R}$}
          \STATE $\bF^{(r)}_k \leftarrow \frac{\sqrt{\frac{P}{\Nrru}}}{\|\bF_k^{(r)}\|_F}\bF^{(r)}_k$
      \ENDFOR  
    \ENDFOR \\
    \RETURN $\left\{ \mathbf{F}_{k} \right\}_{k=1}^K$ and $\left\{ \mathbf{W}_{k} \right\}_{k=1}^K$
  \end{algorithmic}
\end{algorithm}

Unlike the original alternating minimization algorithm in \cite{peters2011cooperative}, which can be shown to converge since leakage power monotonically decreases with each iteration, there is no guarantee that Algorithm \ref{alg:generate_papc_equality_precoders} will in fact converge. The reason is that the projection step $\bF^{(r)}_k \leftarrow \frac{\sqrt{\frac{P}{\Nrru}}}{\|\bF_k^{(r)}\|_F}\bF^{(r)}_k$ may increase the leakage interference objective in some iterations, which destroys the monotonically decreasing sequence used to claim convergence. In some cases, namely for the case of $\Nt = \Nrru$, it is possible to provide a convergent algorithm for IA with per-antenna power constraints. This is achieved by posing each iteration of the alignment process as a least squares problem \cite{yu2010least} over the set of $1\times \Ns$ dimensional subfilters with fixed norm constraints; a number of methods to solve such constrained least squares problem exist \cite{golub2012matrix, bojanczyk1996procrustes, gower2004procrustes}. Since the direct applicability of the least squares formulation is limited to the case of $\Nt = \Nrru$, and since extension to the case of multiple antenna RRUs is non-trivial, we defer the detailed investigation and development of provably convergent algorithms to future work. Despite this shortcoming, Algorithm \ref{alg:generate_papc_equality_precoders} helps demonstrate the practical feasibility of IA with per-RRU power constraints and the numerical results in Section \ref{sec:Simulations} indicate that it tends to converge whenever the properness conditions in Proposition \ref{prop:IAMaxConstraintsProperness} are satisfied.

\section{Simulations} \label{sec:Simulations}
In this section, we first present numerical results demonstrating the feasibility and performance of IA with per-RRU constraints in a baseline Rayleigh fading scenario. This helps verify the properness conditions and power back-off analysis in Sections \ref{sec:MaxConstraints} and \ref{sec:StrictConstraints}. Then we leverage the 3GPP spatial channel model to investigate the potential of IA in DAS in a cluster of seven interfering cells where IA is used to manage inter-cell (intra-cluster) interference.

\subsection{Rayleigh Fading Results}

\begin{figure}[t!]
  \centering
  \includegraphics[width=5.8in]{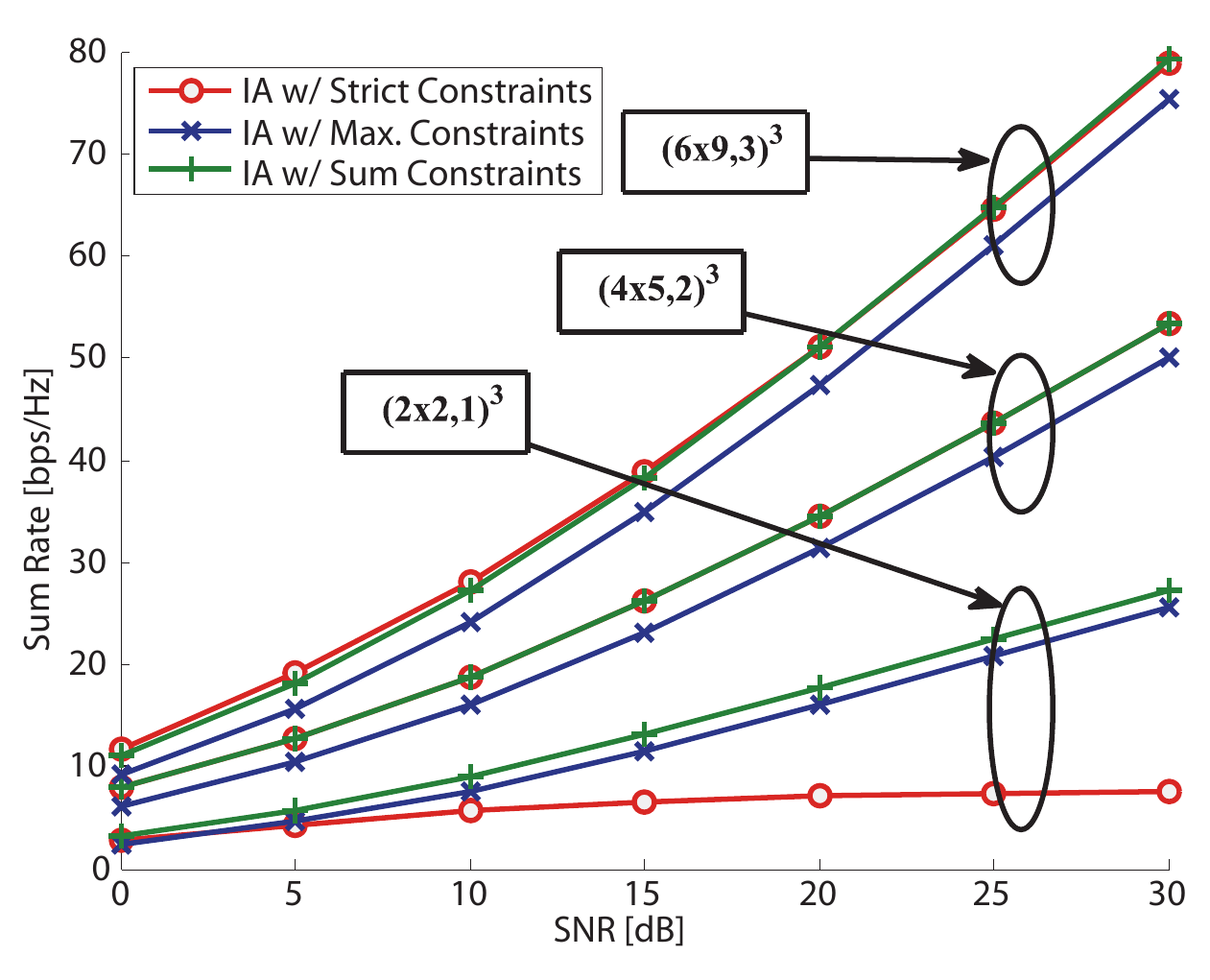}
  \caption{Comparisons of IA with maximum and strict per-antenna power constraints against IA with only total power constraints. The comparisons are made over three different systems labeled using the $(\Nt\times\Nr,\Ns)^K$ shorthand. As we can see from the plots, IA with strict power constraints obtains the same performance as IA with only total power constraints for every system except $(2 \times 2, 1)^3$ in which IA with strict constraints is infeasible, and IA with maximum power constraints suffers from a loss of performance relative to unconstrained IA for every system.}
  \label{fig:papc_inequality_vs_equality}
\end{figure}

In Fig. \ref{fig:papc_inequality_vs_equality}, we compare the average sum-rate performance of (i) traditional IA with only total power constraints, (ii) IA with a maximum antenna power constraint, and (iii) IA with strict per-antenna power constraints. Recall that per-antenna constraints correspond to the case where $\Nt = \Nrru$. For this comparison, we consider the following three IA network configurations
\begin{equation}
(\Nt \times \Nr, \Ns)^K \in \left\{(2 \times 2, 1)^3,\ (4 \times 6, 2)^3,\ (6 \times 9, 3)^3\right\}.
\end{equation}
The power back-off strategy described in Section \ref{sec:MaxConstraints} is used in the case of maximum power constrained
systems, whereas the algorithm in Section \ref{sec:StrictConstraints} is used in the case of strict power constraints. Figure \ref{fig:papc_inequality_vs_equality} shows that, for the two systems $(4 \times 6, 2)^3$ and $(6 \times 9, 3)^3$, Algorithm \ref{alg:generate_papc_equality_precoders} for IA with strict power constraints achieves the same performance as the traditional IA model where only total power is limited. In the case of the $(2 \times 2, 1)^3$ system, however, Fig. \ref{fig:papc_inequality_vs_equality} indicates that Algorithm \ref{alg:generate_papc_equality_precoders} performs poorly and achieves a multiplexing gain of zero. Examining the properness conditions defined in (\ref{eqn:IAMaxConstraintsProperness}), we see that the $(2 \times 2, 1)^3$ system is considered improper and thus IA with per-antenna power constraints was not expected to be feasible in this case. The $(4 \times 6, 2)^3$ and $(6 \times 9, 3)^3$ systems, however, are considered proper. 

\begin{figure}[t!]
  \centering
  \includegraphics[width=5.8in]{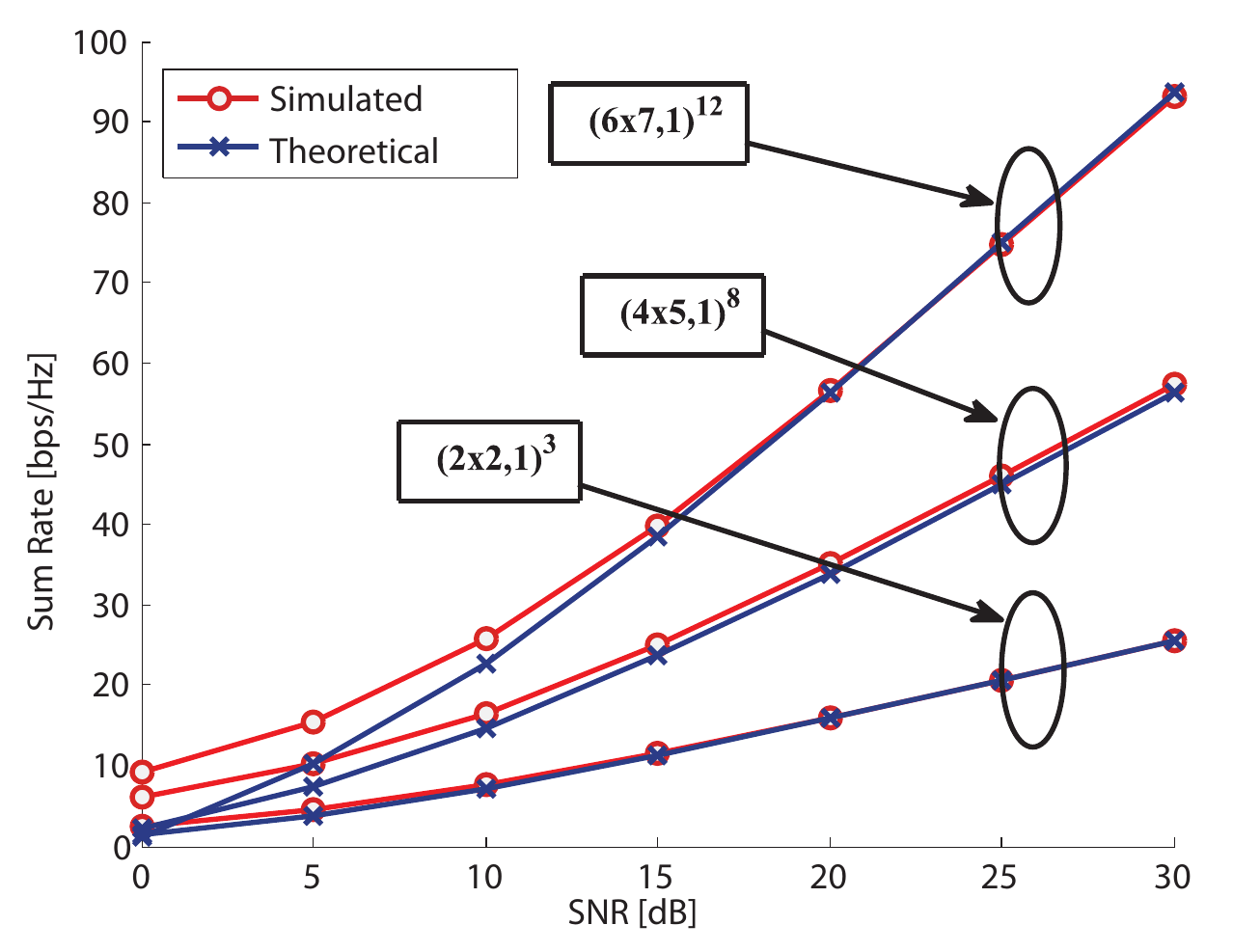}
  \caption{Comparisons of the theoretical versus experimental performances of IA with maximum power constraints with transmit power back-off. These comparisons are made over three different systems labeled using the $(\Nt \times \Nr,\Ns)^K$ shorthand. As we can see, for high SNR, the theoretical performance predicted by the back-off statistics in Section \ref{sec:MaxConstraints} closely matches simulated IA performance.}
  \label{fig:papc_inequality_sum_rate_loss}
\end{figure}

Finally, Fig. \ref{fig:papc_inequality_vs_equality} verifies the fact that satisfying maximal power constraints via the power back-off strategy outlined in Section \ref{sec:MaxConstraints} results in a systematic loss in sum-rate. To investigate the accuracy of the power back-off statistics derived in Section \ref{sec:MaxConstraints}, Fig. \ref{fig:papc_inequality_sum_rate_loss} plots IA's simulated performance as well as the performance predicted by the analysis of Section \ref{sec:MaxConstraints}. To compute IA's predicted performance, the average high-SNR sum-rate loss is evaluated by numerically integrating (\ref{eqn:RateLoss}) using the power back-off distribution in (\ref{eqn:CDFApproximation}). The computed sum rate loss is then subtracted from the average performance of “unconstrained” IA which can be evaluated numerically or using the rate expressions in \cite{elayach2012overhead}. For the three systems $(2 \times 2, 1)^3$, $(4 \times 5, 1)^7$, and $(6 \times 7, 1)^{12}$, Fig. \ref{fig:papc_inequality_sum_rate_loss} shows close agreement between predicted and simulated performance in the high-SNR regime which was considered in our analysis.

\subsection{Spatial Channel Model}

\begin{figure}[t!]
  \subfigure[Centralized antenna system topology]{\includegraphics[width=0.48\linewidth]{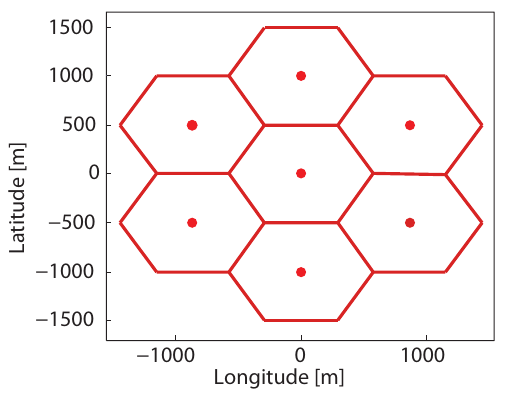}} \hfill
  \subfigure[Distributed antenna system topology]{\includegraphics[width=0.48\linewidth]{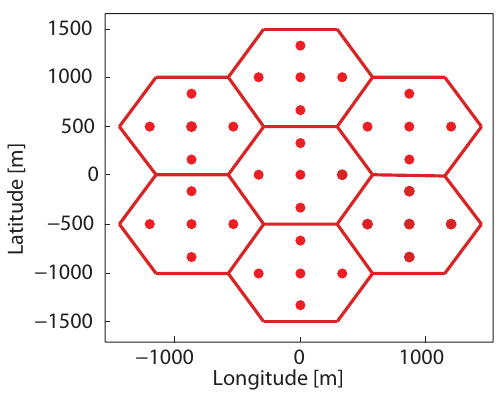}}
  \caption{Network topologies assumed in our simulation study. Nodes correspond to base stations in the co-located antenna systems case and RRUs in the DAS case.}
  \label{fig:network_topologies}
\end{figure}

We now consider the average sum-rate performance of IA in cellular systems with either co-located or distributed antennas. To do so, we leverage the 3GPP spatial channel model to generate frequency-flat channels in the urban-micro scenario with shadowing as described in Section 5 of 3GPP TR 25.996 \cite{3gpp2003scm}. All users' channels are generated using the same spatial channel model parameters. The interested reader is referred to \cite{3gpp2003scm} for additional modeling details concerning path loss, standard antenna patterns and gains, correlated shadowing, and the subscattering implemented in the spatial channel model.

We consider the network topologies given in Figure \ref{fig:network_topologies} in which a cluster of seven base stations each serve one user, and IA is used to manage the inter-cell (intra-cluster) interference. In the traditional system with a central base station, we assume that antennas are placed in a linear array with a 1m inter-element spacing. This corresponds to an inter-element spacing of about 7 wavelengths for a carrier frequency of 2 GHz. In the case of DAS, we denote the cell radius as $R$, and we place a radio unit at the center in addition to four remote radio units at a distance of $2R/3$ from the cell center. Further, we assume that each cell is limited to a total transmit power budget of 46 dBm and that the noise power for each receiver is -106 dBm. When performing IA in systems with co-located antennas, we do not enforce any per-antenna power constraints on top of the standard total power constraint. We note that in practical systems, however, even co-located antennas must adhere to per-antenna power constraints \cite{love2003equal, sesia2009lte}.

\begin{figure}[t!]
  \centering
  \subfigure[IA in Co-located Antenna Systems]{
    \includegraphics[width=0.46\linewidth]{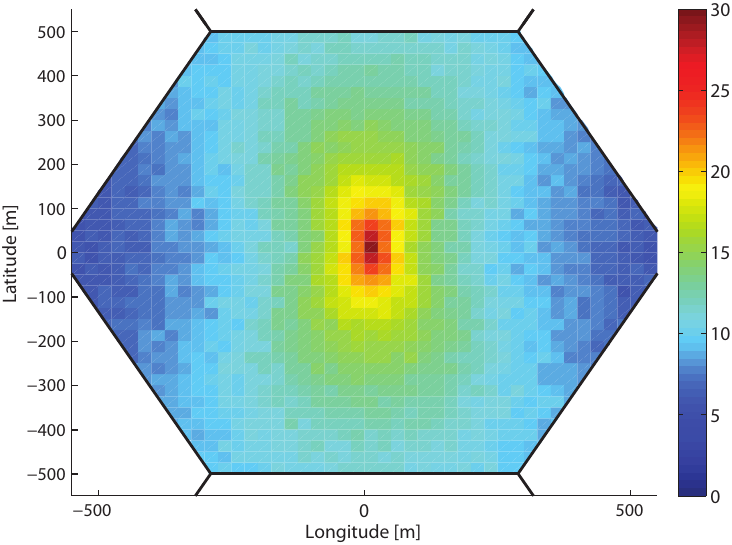}
    \label{subfig:sum_rate:CASIA}
  } \hfill
  \subfigure[IA in Distributed Antenna Systems with Maximum Power Constraints]{
    \includegraphics[width=0.46\linewidth]{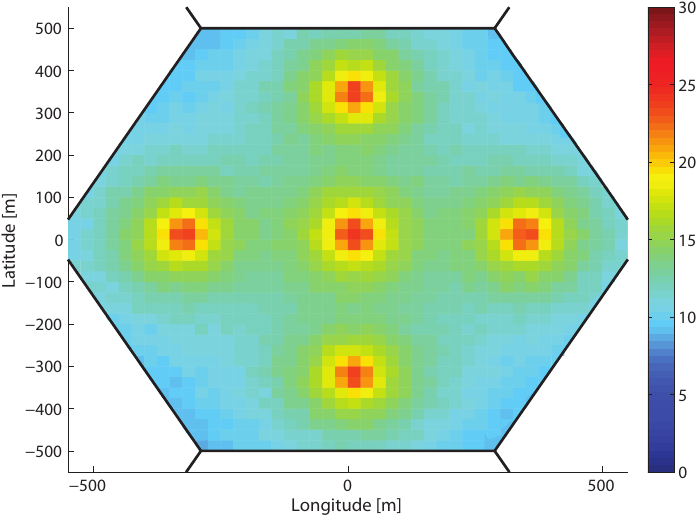}
    \label{subfig:sum_rate:DASIA}
  } \\
  \subfigure[Antenna Selection in Distributed Antenna Systems]{
    \includegraphics[width=0.46\linewidth]{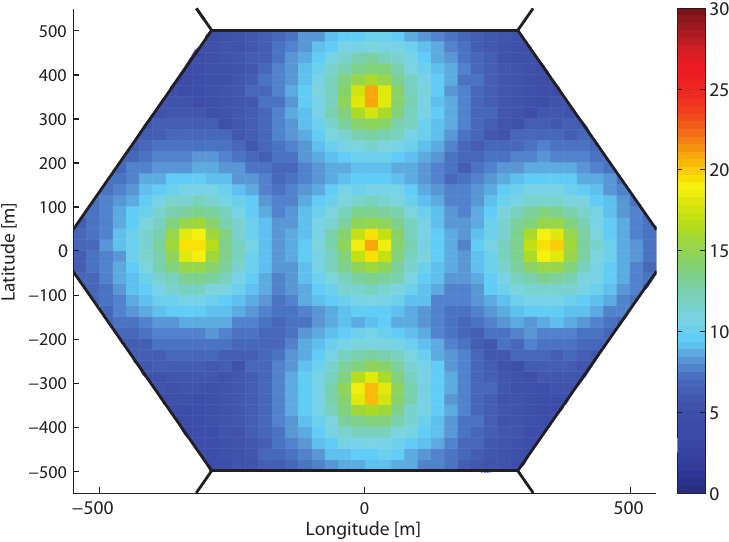}
    \label{subfig:sum_rate:DASAS}
  }
  \caption{The geographical distribution of achievable single-user rate in the center coordinated cell. The cell boundary is given by the black hexagonal overlay. The points of high achievable rate indicate the locations of antennas. IA in DAS is shown to outperform IA with co-located antennas especially at the cell edge. IA is also shown to significantly outperform antenna selection in distributed antenna systems.}
  \label{fig:2DRateDistribution}
\end{figure}

Fig. \ref{fig:2DRateDistribution} plots the data rates achieved by users in the center cell for a single stream system $(\Ns = 1)$ in which each mobile station has $\Nr = 3$ antennas and each base stations has $\Nt = 5$ antennas, either co-located or distributed among $\Nrru = 5$ RRUs. Each data point in Fig. \ref{fig:2DRateDistribution} is generated by averaging over multiple network realizations with randomly generated propagation channels and uniformly distributed mobile user locations in each of the six interfering cells. As we can see from the Fig. \ref{fig:2DRateDistribution}, distributed antennas enable IA to perform uniformly-well throughout the cell. More importantly, applying IA in DAS results in significant cell-edge data rates when compared to the same transmission strategy in traditional cellular systems with a centralized base station. Thus, IA can benefit from the SNR boost provided by DAS to overcome its sub-optimal performance in low-to-medium SNR scenarios. Moreover, Fig. \ref{fig:2DRateDistribution} compares the performance of IA in DAS to the rates achieved by antenna or RRU selection. RRU selection is a simple transmission strategy in which a mobile user is served, via single-user transmission techniques, by the RRU closest to it. RRU selection is among the main transmission strategies typically considered in DAS. From Fig. \ref{fig:2DRateDistribution} we see that IA strictly outperforms RRU selection throughout the entire cell and thus constitutes a promising candidate transmission strategy for DAS.

\begin{figure}[t!]
  \centering
  \includegraphics[width=6in]{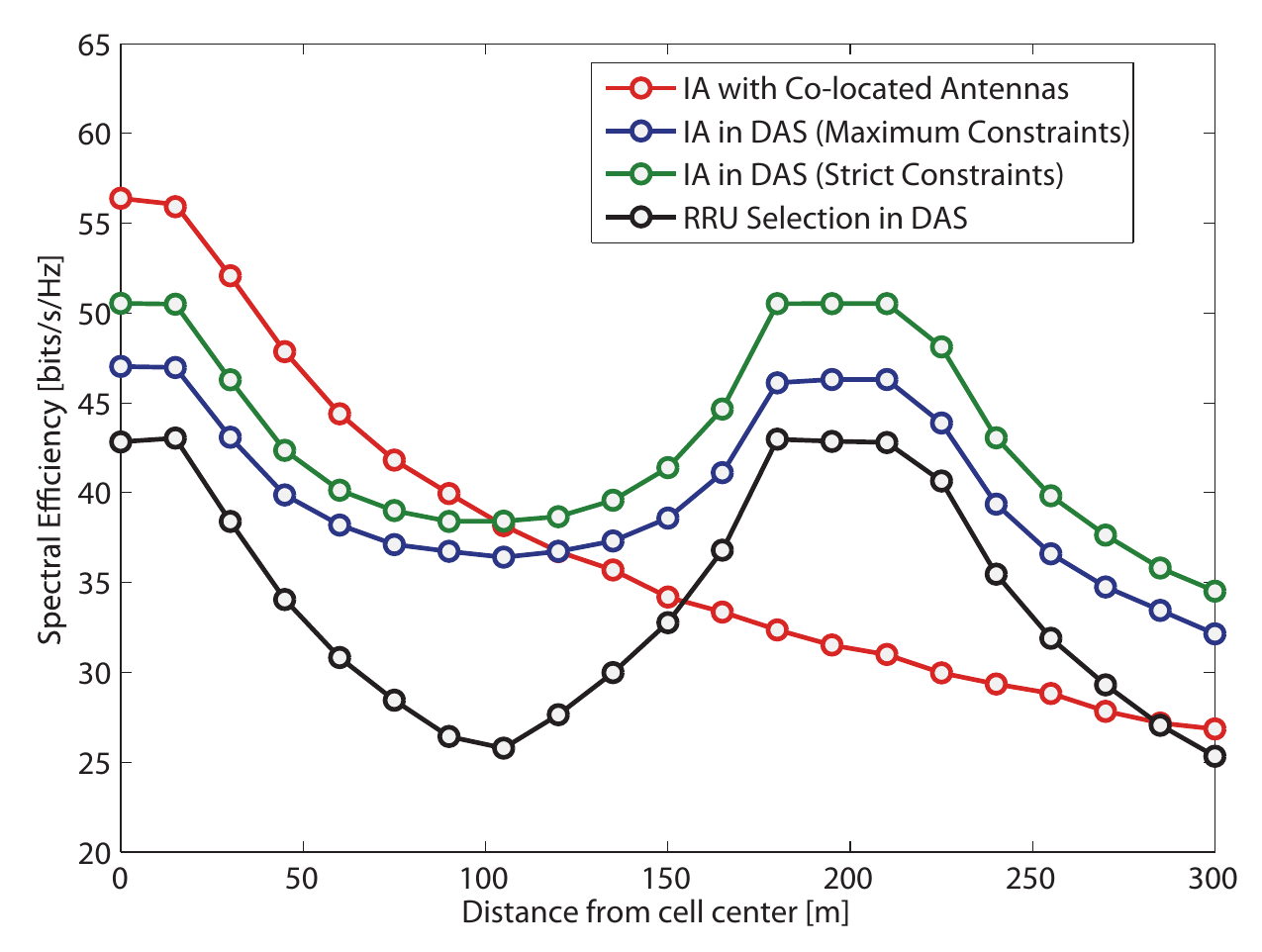}
  \caption{The rates achieved by users in the center cell of a $(15 \times 5, 2)^7$ system with $\Nrru = 5$ RRUs. The plot compares rates achieved by IA in a system with co-located antennas, IA in DAS with either maximum or strict power constraints, or RRU selection. In the case of RRU selection, each RRU transmits 2 data streams to its receiver via single-user precoding and combining.}
  \label{fig:RatesVsDistance}
\end{figure}

To further explore the rates achieved by IA, and to examine the performance of multi-stream transmission with multiple antenna RRUs, Fig. \ref{fig:RatesVsDistance} plots the average data rate achieved by users as a function of their distance from the cell center. The system considered in Fig. \ref{fig:RatesVsDistance} consists of the 5 RRUs with three antennas each (for a total of $\Nt = 15$ transmit antennas). The transmit antennas are used to send $\Ns = 2$ data streams to mobile users with $\Nr = 5$ receiver antennas. Fig. \ref{fig:RatesVsDistance} indicates that IA outperforms RRU selection throughout the entire cell and most significantly in the cell interior. Further, the combination of IA and distributed antennas is shown to provide significant improvements in spectral efficiency at the cell edge. We notice from Fig. \ref{fig:RatesVsDistance} that the distributed antenna IA system with strict per-RRU constraints improves spectral efficiencies for all users further than $100\ m$ away from the cell center. For a cell radius of $300\ m$, this translates into improving the rates of approximately $90\%$ of the mobile users in a cellular system.

\section{Conclusion} \label{sec:Conclusion}
In this paper, we considered the application of IA to distributed antenna systems which may help avoid IA's sub-optimal performance in cellular networks by boosting SNR at the cell-edge. Since the power constraints imposed on DAS are neglected in the existing IA literature, we reevaluated the potential of IA in systems with per-RRU power constraints and proposed simplified methods to compute IA precoders for DAS. For systems in which maximum RRU power is limited, we gave analytical expressions for the distribution of the rate loss caused by per-RRU constraints. For systems in which strict power constraints are enforced, we characterized the set of distributed antenna configurations in which IA is expected to be feasible and provided an iterative IA algorithm for such systems. Using a practical DAS channel model, we showed that the combination of IA and DAS can improve data rates for the majority of mobile users when compared to (i) IA in systems with co-located antennas, and (ii) distributed antennas with existing interference management solutions.


\appendices

\section{Proof of Proposition~\ref{prop:RateLoss}} \label{append:inequality_sum_rate_loss}
\begin{IEEEproof}
At sufficiently \emph{high signal-to-noise ratio}, the interference zero-forcing receive filter $\bW_k\ \forall k \in \mathcal{K}$ becomes rate-optimal and the average sum-rate achievable by the IA solutions in \cite{cadambe2008interference, peters2011cooperative, gomadam2008approaching} is given by
\begin{equation}
\Rsum\approx \mathbb{E}_\mathcal{H}\left[\sum\limits_{k=1}^{K}\log_2\left|\left(\sigma^2 \bI_\Ns+\bJ_k\right)^{-1}\bS_k\right|\right],
\end{equation}
where the approximation is a result of adopting a high-SNR assumption, the matrix $\bJ_k = \sum_{\ell \neq k} \bW_k^* \bH_{k\ell}\bF_\ell\bF_\ell^*\bH_{k\ell}^*\bW_k$ is the interference covariance matrix and $\bS_k =\bW_k^* \bH_{kk}\bF_k\bF_k^*\bH_{kk}^*\bW_k$ is the post-processing desired signal covariance. In the cases of interest where IA is considered feasible, the conditions $\bW_k^* \bH_{k\ell}\bF_\ell = \mathbf{0}_\Ns$ and $\bW_k^*\bW_k = \bI_\Ns$ are satisfied $\forall k\neq \ell$. Thus, $\bJ_k = 0\ \forall k$, and we can write
\begin{equation}
\Rsum\approx \mathbb{E}_\mathcal{H}\left[\sum\limits_{k=1}^{K}\log_2\left|\frac{1}{\sigma^2}\bS_k\right|\right].
\end{equation}
Similarly, the average sum-rate achieved with the maximum power constraint of (\ref{eqn:StrictConstraintsFormulation}), is given by
\begin{equation}
\bar{R}_\mathrm{max-power} \approx \mathbb{E}_\mathcal{H}\left[\sum\limits_{k=1}^{K}\log_2\left|\frac{1}{\sigma^2}\frac{P}{\Nrru\beta^2_k}\bS_k\right|\right].
\end{equation}
Using the scaling determinant rule and the linearity of expectation, 
the loss in sum-rate incurred by enforcing maximum per-RRU power constraints simplifies to
\begin{equation}
\bar{R}_\mathrm{loss}=\Rsum - \bar{R}_\mathrm{max-power} \approx \sum\limits_{k=1}^{K}\Ns \mathbb{E}\left[\log_2\left(\frac{P}{\Nrru\beta^2_k}\right)\right].
\end{equation}
\end{IEEEproof}

\section{Proof of Proposition~\ref{prop:CDFApproximation}} \label{append:inequality_max_row_norm_cdf_beamforming}
\begin{IEEEproof}
Recall that the maximum per-RRU transmit power is defined as
\begin{displaymath}
    \beta^{2} = \max_{r \in \rruset} \|\mathbf{f}^{(r)}\|_2^{2}.
\end{displaymath}
where $\mathbf{f}^{(r)}$ for $r \in \rruset$ is the subvector in $\mathbf{f}$ corresponding to the beamforming weights used by the antennas in the $r^{th}$ \ac{RRU}. Also recall that generating a random vector from the uniform distribution over the $\Nt$-dimensional hypersphere is equivalent to performing the following steps~\cite{tulino2004random}: (i) randomly generate $\mathbf{x} \sim \mathcal{CN}(\mathbf{0}, \bI_{\Nt})$, then (ii) calculate $\mathbf{f}$ as $\mathbf{f} = \mathbf{x} / \norm{\mathbf{x}}_{2}$. Thus, we have that
\begin{displaymath}
  \beta^{2} = \max_{r \in \rruset}{\norm{\mathbf{f}^{(r)}}_{2}^{2}} 
    \stackrel{(d)}{=} \max_{r \in \rruset} \frac{\norm{\mathbf{x}^{(r)}}_{2}^{2}}{\norm{\mathbf{x}}_{2}^{2}}.
\end{displaymath}
where $\mathbf{x}^{(r)}$ is the subvector of $\mathbf{x}$ in the same manner that $\mathbf{f}^{(r)}$ is the subvector of $\mathbf{f}$.
From this, we can write
\begin{displaymath}
  \mathbb{P}_{\beta^{2}}\left\{ \beta^{2} \leq \frac{x}{N_{t}} \right\} =
    \mathbb{P}_{\mathbf{x}}\left\{ \max_{r \in \rruset} \frac{\norm{\mathbf{x}^{(r)}}_{2}^{2}}{\norm{\mathbf{x}}_{2}^{2}} \leq \frac{x}{N_{t}} \right\}.
\end{displaymath}
By the strong law of large numbers, note that
\begin{displaymath}
  \frac{1}{N_{t}} \norm{\mathbf{x}}_{2}^{2} \xrightarrow{(\text{a.s.})} \mathbb{E}\left[ \abs{x_{1}}^{2} \right] = 1 \text{ as } N_{t} \rightarrow \infty,
\end{displaymath}
where $x_1$ is the first element of $\bx$ and $\mathbb{E}\left[ \abs{x_{1}}^{2} \right] = 1$  by definition.  Therefore, for all $\epsilon > 0$ and all $N_{t}$ sufficiently large, we have that
\begin{displaymath}
  \mathbb{P}_{\mathbf{x}}\left\{ \abs{ \frac{1}{N_{t}} \norm{\mathbf{x}}_{2}^{2} - 1 } < \epsilon \right\} = 1
\end{displaymath}
which using the equivalence $\abs{x - 1} < \epsilon \iff 1 - \epsilon < x < 1 + \epsilon$, this can be rewritten as
\begin{displaymath}
  \mathbb{P}_{\mathbf{x}}\left\{ 1 - \epsilon < \frac{1}{N_{t}} \norm{\mathbf{x}}_{2}^{2} < 1 + \epsilon \right\} = 1.
\end{displaymath}
Using the above high-probability bounds for $\norm{\mathbf{x}}_{2}^{2}$ and the fact the \ac{CDF} is right-continuous and non-decreasing, we can bound $\mathbb{P}_{\beta^{2}}\left\{ \beta^{2} \leq \frac{x}{N_{t}} \right\}$ as follows:
\begin{align*}
  \mathbb{P}_{\mathbf{x}} & \left\{ \max_{n \in \{1, \dotsc, N_{t}\}} \frac{{\abs{x_{n}}^{2}}}{1 - \epsilon} \leq x \right\}
    < \mathbb{P}_{\beta^{2}}\left\{ \beta^{2} \leq \frac{x}{N_{t}} \right\}
    < \mathbb{P}_{\mathbf{x}} \left\{ \max_{n \in \{1, \dotsc, N_{t}\}} \frac{{\abs{x_{n}}^{2}}}{1 + \epsilon} \leq x \right\}.
\end{align*}
Using the fact that the norm of $\mathbf{x}^{(r)}$ for $r \in \rruset$ is independent and chi-squared distributed with $N_{t} / N_{RRU}$ degrees of freedom, we have
\begin{align*}
  \chisqrcdf{ (1 - \epsilon)x }{\frac{N_{t}}{N_{RRU}}}^{N_{t}}
    < \mathbb{P}_{\beta^{2}}\left\{ \beta^{2} \leq \frac{x}{N_{t}} \right\}
    <  \chisqrcdf{(1 + \epsilon)x}{\frac{N_{t}}{N_{RRU}}}^{N_{t}}.
\end{align*}
Taking the $\log$ of all sides and dividing by $N_{t}$, we have that
\begin{align*}
  \log{\chisqrcdf{(1 - \epsilon)x}{\frac{N_{t}}{N_{RRU}}}}
    < \frac{1}{N_{t}} \log{\mathbb{P}_{\beta^{2}}\left\{ \beta^{2} \leq \frac{x}{N_{t}} \right\}}
    < \log{\chisqrcdf{(1 + \epsilon)x}{\frac{N_{t}}{N_{RRU}}}}.
\end{align*}
Because $\epsilon > 0$ was arbitrary and $\chisqrcdf{\cdot}{\cdot}$ is a continuous, non-decreasing function, it follows that
\begin{displaymath}
   \lim_{N_{t} \to \infty} \frac{1}{N_{t}} \log{\mathbb{P}_{\beta^{2}}\left\{ \beta^{2} \leq \frac{x}{N_{t}} \right\}} = 
       \log{\mathcal{Q}\left( x; \frac{N_{t}}{N_{RRU}} \right)},
\end{displaymath}
which is what we wanted to show.
\end{IEEEproof}

\section{Proof of Proposition~\ref{prop:NumConstraints}} \label{append:papc_equality_properness}
To prove Proposition \ref{prop:NumConstraints}, we will require the following result.
\begin{proposition}
If $\bA$ is a real, square, and diagonally-dominant matrix, then the there exists a
real and non-negative solution $\bd$ to the equation $\bA\bd = \gamma \mathbf{1}$ for any $\gamma \in \mathbb{R}$ such that $\gamma > 0$.
\label{prop:Farkas}
\end{proposition}
\begin{IEEEproof}
The proposition follows immediately from Farkas' lemma \cite{scowcroft2006nonnegative}.
\end{IEEEproof}
Using this proposition, we now proceed to prove the result in Proposition \ref{prop:NumConstraints}. We note that for the case of $\Nrru = \Nt$ a simplified proof is given \cite{starr2011interference} and the proof given in this paper is a generalization to the case $\Nrru \leq \Nt$ which follows in the general footsteps of \cite{starr2011interference}.
\begin{IEEEproof}
To prove the result in Proposition \ref{prop:NumConstraints}, similarly to our result in \cite{starr2011interference}, we show that out of each user's $\Nrru$ sub-filters $\bF^{(r)}_k$ , $\Ns$ of those transmit sub-filters can be designed such that (\ref{eqn:StrictConstraintsFormulation}) is satisfied arbitrarily without affecting the feasibility of alignment. This effectively removes $\Ns$ of each user's $\Nrru$ strict power constraints from the set of unsatisfied equations in (\ref{eqn:StrictConstraintsFormulation}). As a result, the number of non-trivially satisfied equations introduced by (\ref{eqn:StrictConstraintsFormulation}) is equal to $K(\Nrru - \Ns)$ whenever $\Nrru \geq \Ns$.

To show this, let the matrix $\bF_k$ be an arbitrary unconstrained IA precoder. Now consider a partitioning of each subfilter $\bF^{(r)}_k$ between its top row $\left[\bF^{(r)}_k\right]_{1,:}$ and its remaining lower rows $\left[\bF^{(r)}_k\right]_{2:\frac{\Nt}{\Nrru},:}$, i.e, we simply expand $\bF^{(r)}_k$ and write it as $\bF^{(r)}_k=\left[\left[\bF^{(r)}_k\right]_{1,:}^*,\ \left[\bF^{(r)}_k\right]_{2:\frac{\Nt}{\Nrru},:}^*\right]^*$. Further,
consider a notational remapping (permutation) of each user $k$'s precoder $\bF_k$ in which the first row of the first $\Ns$ transmit subfilters are moved to the top rows of $\bF_k$, i.e., define the permuted precoders $\widetilde{\bF}_k$ given by
\begin{equation}
\widetilde{\bF}_k=\left[\left[\bF^{(1)}_k\right]_{1,:}^*,\ \hdots,\ \left[\bF^{(\Ns)}_k\right]_{1,:}^*, \left[\bF^{(1)}_k\right]_{2:\frac{\Nt}{\Nrru},:}^*,\ \hdots,\ \left[\bF^{(\Ns)}_k\right]_{2:\frac{\Nt}{\Nrru},:}^*, \bF_k^{(\Ns+1)*}, \hdots,\ \bF_k^{(\Nrru)*} \right]^*
\end{equation}
Note that this remapping is purely notational and does not affect the structure of the alignment problem in any sense. Finally, given the remapped precoder, let $\bM_k$ and $\bN_k$ be partitions of the newly remapped $\widetilde{\bF}_k$ such that
\begin{align}
\bM_k & = \left[\left[\bF^{(1)}_k\right]_{1,:}^*,\ \hdots,\ \left[\bF^{(\Ns)}_k\right]_{1,:}^*\right]^*, \\
\bN_k & =\left[\left[\bF^{(1)}_k\right]_{2:\frac{\Nt}{\Nrru},:}^*,\ \hdots,\ \left[\bF^{(\Ns)}_k\right]_{2:\frac{\Nt}{\Nrru},:}^*, \bF_k^{(\Ns+1)*}, \hdots,\ \bF_k^{(\Nrru)*}\right]^*.
\end{align}
Thus, we now can write the remapped precoder $\widetilde{\bF}_k = \left[\bM_k^*, \bN_k^*\right]^*$. Note that $\bM_k$ is an $\Ns \times \Ns$ square matrix consisting of the first row of the first $\Ns$ transmit subfilters and $\bN_k$ is of size $(\Nt - \Ns)\times \Ns$ and consists of the remaining $(\Nt - \Ns)$ rows.

As outlined in \cite{yetis2010feasibility, starr2011interference}, each precoder $\bF_k$ (and similarly the remapped precoders $\widetilde{\bF}_k$) contains
$\Ns^2$ extraneous variables that do not affect its span and thus do not act as free design variables, i.e., each precoder can be reduced to its minimal basis representation without affecting alignment. To reduce the remapped precoders $\widetilde{\bF}_k$ to their minimal basis representation, we note that $\bM_k$ is an $\Ns\times \Ns$ matrix and thus composed of $\Ns^2$ elements. Therefore, similarly to \cite{yetis2010feasibility}\cite{starr2011interference} we can limit our attention precoders in which $\bM_k = \sqrt{c}\bI_\Ns$ and are thus of the form $\left[\sqrt{c}\bI_\Ns^*,\ \bN_k^*\right]^*$ where $\sqrt{c}$ is an arbitrary constant of our choosing.

Using the remapped precoders $\widetilde{\bF}_k$, which we have shown can always be written in the form
$\widetilde{\bF}_k = \left[\sqrt{c}\bI_\Ns^*,\ \bN_k^*\right]^*$, we now show that it is possible to construct a real diagonal matrix $\bD_k \succ \mathbf{0}$ such that $\|\bF^{(r)}_k\bD^{1/2}_k\|_F^2=\frac{P}{\Nrru}$ for all $r \in \mathcal{R}$, i.e., such that the first $\Ns$ RRUs automatically satisfy their power constraint. Note that because such post-multiplication does not change the span of a matrix, if the set of precoders $\bF_k\ \forall k\in \mathcal{K}$ achieves alignment, then so will $\bF_k\bD_k^{1/2}$. So, by constructing and post-multiplying by $\bD_k^{1/2}$, we can satisfy the IA conditions while at the same time arbitrarily satisfying the first $\Ns$ equations given in (\ref{eqn:StrictConstraintsFormulation}). We proceed henceforth by constructing such a $\bD_k$.

Let the $\Ns\times \Ns$ matrix $\bA_k$ be constructed such that the element of $\bA_k$ in the $m^\mathrm{th}$ row and $n^\mathrm{th}$ column corresponds to the power applied to the $n^\mathrm{th}$ stream by the $m^\mathrm{th}$ RRU, for $m= 1,\ \hdots,\ \Ns$. As a result, the elements of matrix $\bA_k$ are given by
\begin{equation}
\bA_k(m, n) = \underbrace{c \times \delta(m - n)}_{\text{from first antenna of } m^\mathrm{th} \text{ RRU}} + \underbrace{\sum\limits_{\ell=2}^{\frac{\Nt}{\Nrru}}\left|\bF_k^{(m)}(\ell,n)\right|^2,}_{\text{from remaining antennas of } m^\mathrm{th} \text{ RRU}}\ \forall m, n \in \left\{1,\ 2,\ \hdots,\Ns\right\},
\end{equation}
where we have explicitly broken down the elements of $\bA_k$ into two terms as a result of our construction of $\widetilde{\bF}_k = \left[\sqrt{c}\bI_\Ns^*,\ \bN_k^*\right]^*$ in which the first antenna of the $m^\mathrm{th}$ RRU sends only the $m^\mathrm{th}$ stream scaled by $\sqrt{c}$. Note that since the positive scalar $c$ is arbitrary, it can be chosen large enough such that $\bA_k$ is a diagonally-dominant matrix.

Consider now the vector equation $\bA_k\bd_k =\frac{P}{\Nrru}\mathbf{1}$, where $\mathbf{1}$ denotes the all-ones vector of appropriate dimension. If a solution of the form $\bd_k \succ \mathbf{0}$ exists, then recalling the definition of $\bA_k$ we notice that in this case $\bd_k$ represents an adjustment of the power allocated to each stream by the first $\Ns$ RRUs such that they all automatically satisfy their power constraints. Because $\bA_k$ is diagonally-dominant, however, Proposition \ref{prop:Farkas} indicates that a solution $\bd_k \succ \mathbf{0}$ does in fact always exist, i.e., in our IA system an adjusted power allocation can always be found such that the first $\Ns$ RRUs satisfy their power constraints arbitrarily. So, defining $\bD_k = \mathrm{diag}(\bd_k)$, and using the precoders $\bF_k\bD_k^{1/2}$ we would have satisfied both the alignment conditions in (\ref{eqn:ia_formulation:002}) and the first $\Ns$ of the $\Nrru$ power constraints in (\ref{eqn:StrictConstraintsFormulation}) automatically. Therefore, we see that for each user, (\ref{eqn:StrictConstraintsFormulation}) contains $\Nrru - \Ns$ non-trivially satisfied equations that must still be accounted for whenever $\Nrru \geq \Ns$. Therefore, the total number of non-trivially satisfied equations $N^{(2)}_\mathrm{e}$ contributed by (\ref{eqn:StrictConstraintsFormulation}) is given by
\begin{equation}
N^{(2)}_\mathrm{e} = \max\left\{K (\Nrru - \Ns) , 0\right\}.
\end{equation}
\end{IEEEproof}

\singlespace
\bibliographystyle{IEEEtran}
\bibliography{IEEEabrv,WSILabrv,IA_DAS}

\end{document}